\documentclass[aip,pop,reprint]{revtex4-1}
\usepackage{graphicx}
\usepackage{subfigure}
\usepackage{amsmath}
\usepackage{overpic}
\usepackage{color}
\usepackage{rotating}
\usepackage{txfonts}
\usepackage{array}
\usepackage{multirow}
\usepackage{url}
\draft 

\begin{document}

\title{High quality electron beam acceleration by ionization injection in laser wakefields with mid-infrared dual-color lasers} 

\author{Ming Zeng}
\email{ming.zeng@eli-np.ro}
\affiliation{Key Laboratory for Laser Plasmas (Ministry of Education), Department of Physics and Astronomy, Shanghai Jiao Tong University, Shanghai 200240, China}
\affiliation{Collaborative Innovation Center of IFSA (CICIFSA), Shanghai Jiao Tong University, Shanghai, 200240, China}
\affiliation{Extreme Light Infrastructure - Nuclear Physics, Horia Hulubei National Institute for Physics and Nuclear Engineering, 30 Reactorului Street, P.O. Box MG-6, 077125 Magurele, jud. Ilfov, Romania}
\author{Ji Luo}
\affiliation{Key Laboratory for Laser Plasmas (Ministry of Education), Department of Physics and Astronomy, Shanghai Jiao Tong University, Shanghai 200240, China}
\affiliation{Collaborative Innovation Center of IFSA (CICIFSA), Shanghai Jiao Tong University, Shanghai, 200240, China}
\author{Min Chen}
\affiliation{Key Laboratory for Laser Plasmas (Ministry of Education), Department of Physics and Astronomy, Shanghai Jiao Tong University, Shanghai 200240, China}
\affiliation{Collaborative Innovation Center of IFSA (CICIFSA), Shanghai Jiao Tong University, Shanghai, 200240, China}
\author{Warren B. Mori}
\affiliation{University of California, Los Angeles, California 90095, USA}
\author{Zheng-Ming Sheng}
\affiliation{Key Laboratory for Laser Plasmas (Ministry of Education), Department of Physics and Astronomy, Shanghai Jiao Tong University, Shanghai 200240, China}
\affiliation{Collaborative Innovation Center of IFSA (CICIFSA), Shanghai Jiao Tong University, Shanghai, 200240, China}
\affiliation{SUPA, Department of Physics, University of Strathclyde, Glasgow G4 0NG, UK}
\author{Bernhard Hidding}
\affiliation{SUPA, Department of Physics, University of Strathclyde, Glasgow G4 0NG, UK}

\begin{abstract}
For the laser wakefield acceleration, suppression of beam energy spread while keeping sufficient charge is one of the key challenges. In order to achieve this, we propose bichromatic laser ionization injection with combined laser wavelengths of $2.4\rm \mu m$ and $0.8\rm \mu m$ for wakefield excitation and for triggering electron injection via field ionization, respectively. A laser pulse at $2.4\rm \mu m$ wavelength enables one to drive an intense acceleration structure with relatively low laser power. To further reduce the requirement of laser power, we also propose to use carbon dioxide as the working gas medium, where carbon acts as the injection element. Our full three dimensional particle-in-cell simulations show that electron beams at the GeV energy level with both low energy spreads (around one percent) and high charges (several tens of picocoulomb) can be obtained by this scheme with laser parameters achievable  in the near future.
\end{abstract}

\maketitle
Ever since its invention, the laser wakefield accelerator (LWFA) has been considered as one of the most promising candidates of the next generation of accelerators~\cite{TajimaPRL1979}. Compared with conventional radio-frequency accelerators, a laser wakefield accelerator has the advantage of several orders higher acceleration gradient, meanwhile currently it has the drawbacks of relatively poor beam qualities. Great progresses has been made over the past years~\cite{ManglesNature2004, GeddesNature2004, FaureNature2004, EsareyRMP2009, MalkaPOP2012, LeemansPRL2014}. Nevertheless, to further improve the beam quality including the charge, peak energy, energy spread, and emittance is still one of the top priorities in the community in order to make the LWFA suitable for applications.

There are mainly two ways of improving the output beam quality. One is improving the beam phase space distribution during the acceleration processes~\cite{GuillaumePRL2015}, and the other is improving the injection processes at the very beginning of the acceleration~\cite{BuckPRL2013, LehePRL2013, VieiraPRL2011}. Among the variety of injection schemes, the ionization-induced injection is found to be simple and effective~\cite{UmstadterPRL1996, MinChenJAP2006, OzPRL2007, McGuffeyPRL2010, PakPRL2010, ClaytonPRL2010, PollockPRL2011, JSLiuPRL2011, MChenPOP2012}. By using different variations of this mechanism, electron beams with low emittances down to the nano-meter level~\cite{HiddingPRL2012, XLXuPRL2014, LLYuPRL2014}, or low energy spreads down to a few percent~\cite{MZengJPP2012, MZengPOP2014, MMirzaieSRep2015} were produced. Recently, a new ionization injection variation utilizing the beating of bichromatic lasers to produce sub-percent energy spread is proposed~\cite{MZengPRL2015}. In this scheme, the driver is a femtosecond laser pulse with two frequency components $\omega_{1, 2}$, and the laser peak electric field amplitude evolves due to the dispersion difference of the two frequency components. The evolution length period of the electric field amplitude is
\begin{eqnarray}\label{eq:period_z}
  \Delta z = \dfrac{4\pi c \omega_1}{\omega_p^2(\dfrac{\omega_2}{\omega_1}-\dfrac{\omega_1}{\omega_2})},
\end{eqnarray}
which is typically in hundred micrometers or millimeter scales, where $\omega_p$ is the plasma frequency. Consequently, the ionization triggered injections only occur in comparably confined volumes. The optimal combination of bichromatic laser components is found to have the ratio of $\omega_1 / \omega_2 = \frac{1}{3}$ and $E_{10} / E_{20} = 3$, where $E_{10}$ and $E_{20}$ are the electric field amplitude of the two components. Using this certain combination, the electric field wave form of the laser switches between $\sin(\omega t)+\frac{1}{3}\sin(3\omega t)$ and $\cos(\omega t)+\frac{1}{3}\cos(3\omega t)$ when propagating in the plasma, and resembles a square wave in the former form. Thus this scheme is also called the square-wave-like bichromatic laser (SWBL) injections.

\begin{figure}
  \centering
  \begin{tabular}{cc}
    \subfigure{\label{fig:ion_single}
    \begin{overpic}[width=0.24\textwidth]{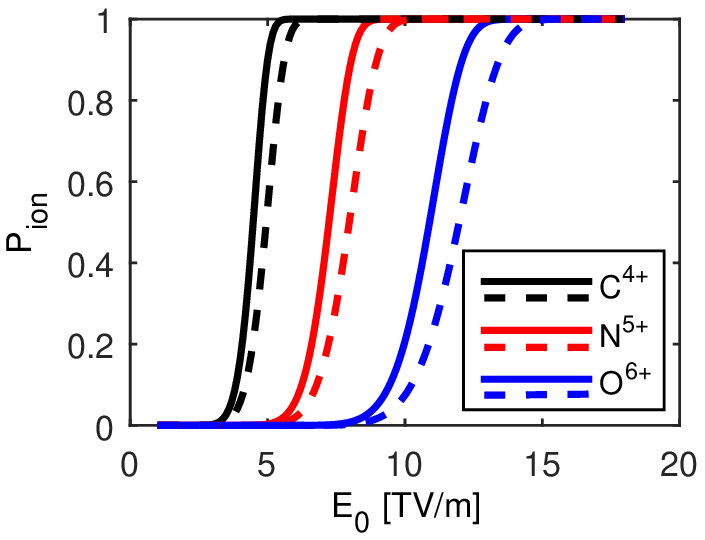}
      \put(20,60){(a)}
    \end{overpic}} &
    \subfigure{\label{fig:ion_bichrom}
    \begin{overpic}[width=0.24\textwidth, trim=20 0 -20 0]{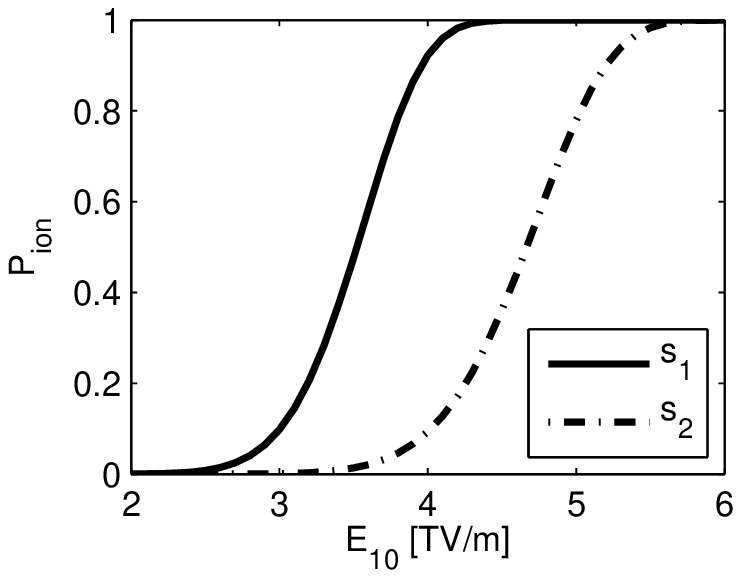}
      \put(12,60){(b)}
    \end{overpic}}
  \end{tabular}
  \caption{\label{fig:ion_prob} Ionization probability after one laser cycle vs.\ laser electric field amplitude for different situations predicted by the ADK model. (a) The laser is either with the wavelength of $2.4\ \rm \mu m$ (solid lines) or  with $0.8\ \rm \mu m$ (dash lines). The black/red/blue colors represent the cases of $\rm C^{4+}$,  $\rm N^{5+}$ and $\rm O^{6+}$, whose L-shells have already been stripped off by laser pre-pulses. (b) The laser is square-wave-like bichromatic with wavelengths of $2.4\ \rm \mu m$ and $0.8\ \rm \mu m$, and the ionization object is $\rm C^{4+}$. The solid line shows the ionization probability after one laser cycle when the combined electric field takes the form of $E_{10}\left[\cos(\omega t)+\frac{1}{3}\cos(3\omega t)\right]$, and the dash-dot line shows that when it takes the form of $E_{10}\left[\sin(\omega t)+\frac{1}{3}\sin(3\omega t)\right]$. }
\end{figure}

In this letter, we extend the SWBL injection scheme to the mid-infrared laser region. Instead of a 800 nm laser combined with a frequency tripled split off part~\cite{MZengPRL2015}, we use a laser pulse with 2.4 $\rm \mu m$ wave length and combined with a 800 nm laser. Such 2.4 $\rm \mu m$ laser pulse in the 100 TW level using OPCPA technique is already under design~\cite{FYWangLPL2015}. Moreover, we choose a few-cycle 800 nm laser pulse as the $\omega_2$ component, which is a standard laser technique~\cite{VaupelOpEng2013, HerrmannOL2009, TavellaOE2006}. Carbon dioxide is chosen as the injection gas, but the actual injection element is carbon instead of oxygen. This reduces the required laser intensity to less than a half compared with the case of using nitrogen.


Many of the existing studies use nitrogen as the injection gas, because the ionization threshold of the nitrogen inner shell is very close to the femtosecond laser electric field amplitude that widely achievable recently~\cite{PakPRL2010, PollockPRL2011}. There are also a few use carbon dioxide or oxygen as the injection gas, but only oxygen atoms contribute to the injections~\cite{ClaytonPRL2010, JSLiuPRL2011}. In order to see the different ionization thresholds for the inner shell of a few such elements, we use the widely accepted ADK model~\cite{AmmosovJETP1986, BruhwilerPOP2003, MChenJCP2013}. According to this model, the ionization probabilities for different elements under different laser field amplitudes are shown in Fig.~\ref{fig:ion_prob}.

Figure~\ref{fig:ion_single} shows the ionization probability after one laser cycle vs.\ electric field amplitude either for $2.4\ \rm \mu m$  or for  $0.8\ \rm \mu m$ lasers. Only the first K-shell electrons are considered. We do not consider the L-shell electrons because they can be fully ionized by the electric fields which are even one magnitude smaller than the one used in the plot. And we also do not plot the ionization probability for the second K-shell electrons because they are not necessary in our discussions. We can see that for notable ionization probabilities ($P_{\rm ion} \gtrsim 1\%$), the electric field amplitude should exceed about 3.3, 5.4 and 8.1 TV/m for $\rm C^{4+}$,  $\rm N^{5+}$, and $\rm O^{6+}$, respectively. Thus the ratio of the laser intensity thresholds to ionize the inner shell of these three elements is about 1:2.7:6. In many simulations, we empirically find that a relatively larger laser spot size (thus lower laser intensity if the laser power is fixed) and lower plasma density can increase both the final output electron beam energy and charge. So we choose carbon as the ionization injection element in the following discussions.

Figure~\ref{fig:ion_bichrom} shows the ionization probability after one laser cycle when a combined SWBL laser is used. The solid and dash-dot lines show the ionization probability when the laser is in the form of $E_{10}\left[\cos(\omega t)+\frac{1}{3}\cos(3\omega t)\right]$ and $E_{10}\left[\sin(\omega t)+\frac{1}{3}\sin(3\omega t)\right]$, respectively. The big difference of these two lines is because that the ionization rate grows exponentially when the peak electric field increases~\cite{MChenJCP2013}. One can see that the thresholds for notable ionization probabilities are 2.5 and 3.4 TV/m for the two waveforms, respectively. It means that the intermittent K-shell ionization can occur if $E_{10}$ is between 2.5 and 3.4 TV/m.

\begin{figure}
  \centering
  \includegraphics[width=0.48\textwidth]{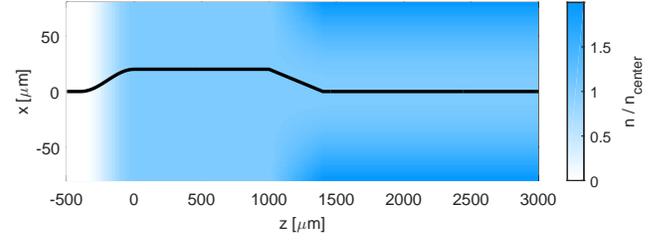}
  \caption{\label{fig:channel} Schematic diagram of the plasma profile in the simulations. The pseudo-color plot shows the distribution of the background plasma electrons, and the black solid line in the middle shows the longitudinal density profile of $\rm C^{4+}$. The region from $z=0$ to $1000\ \rm \mu m$ is uniform for both the background plasma and $\rm C^{4+}$. The region from $z=1400\ \rm \mu m$ to infinite is a transversely parabolic channel without $\rm C^{4+}$. The region from $z=-400$ to $0\ \rm \mu m$ is a transition from vacuum to the plasma, and the region from $z=1000$ to $1400\ \rm \mu m$ is a transition from the uniform region to the channel. The axial plasma density in the channel is the same as the density of the uniform region.}
\end{figure}

\begin{figure*}
  \centering
  \begin{tabular}{llll}
    \subfigure{\label{fig:snapshot2}
    \begin{overpic}[width=0.21\textwidth]{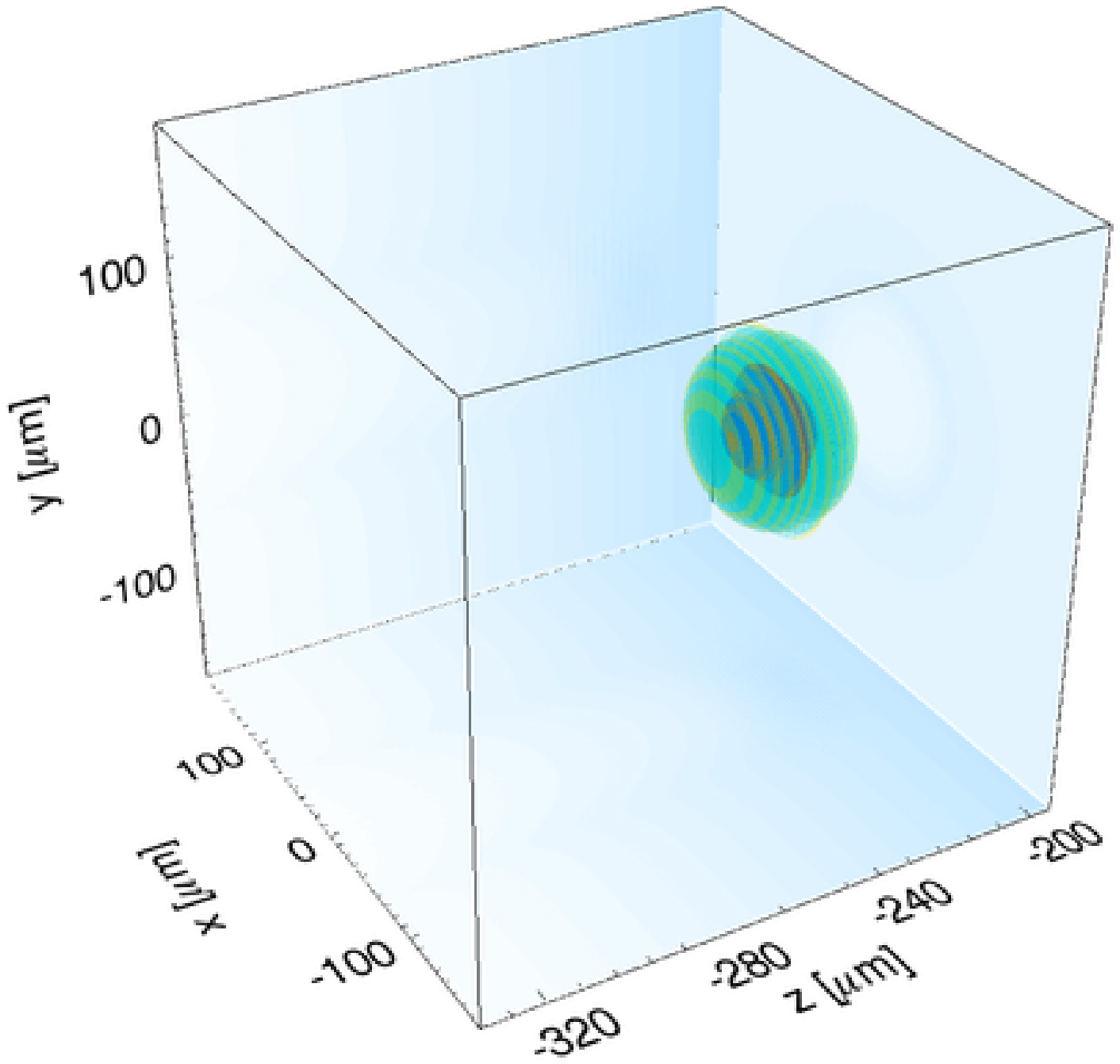}
      \put(24,78){(a)}
    \end{overpic}} &
    \subfigure{\label{fig:snapshot9}
    \begin{overpic}[width=0.21\textwidth]{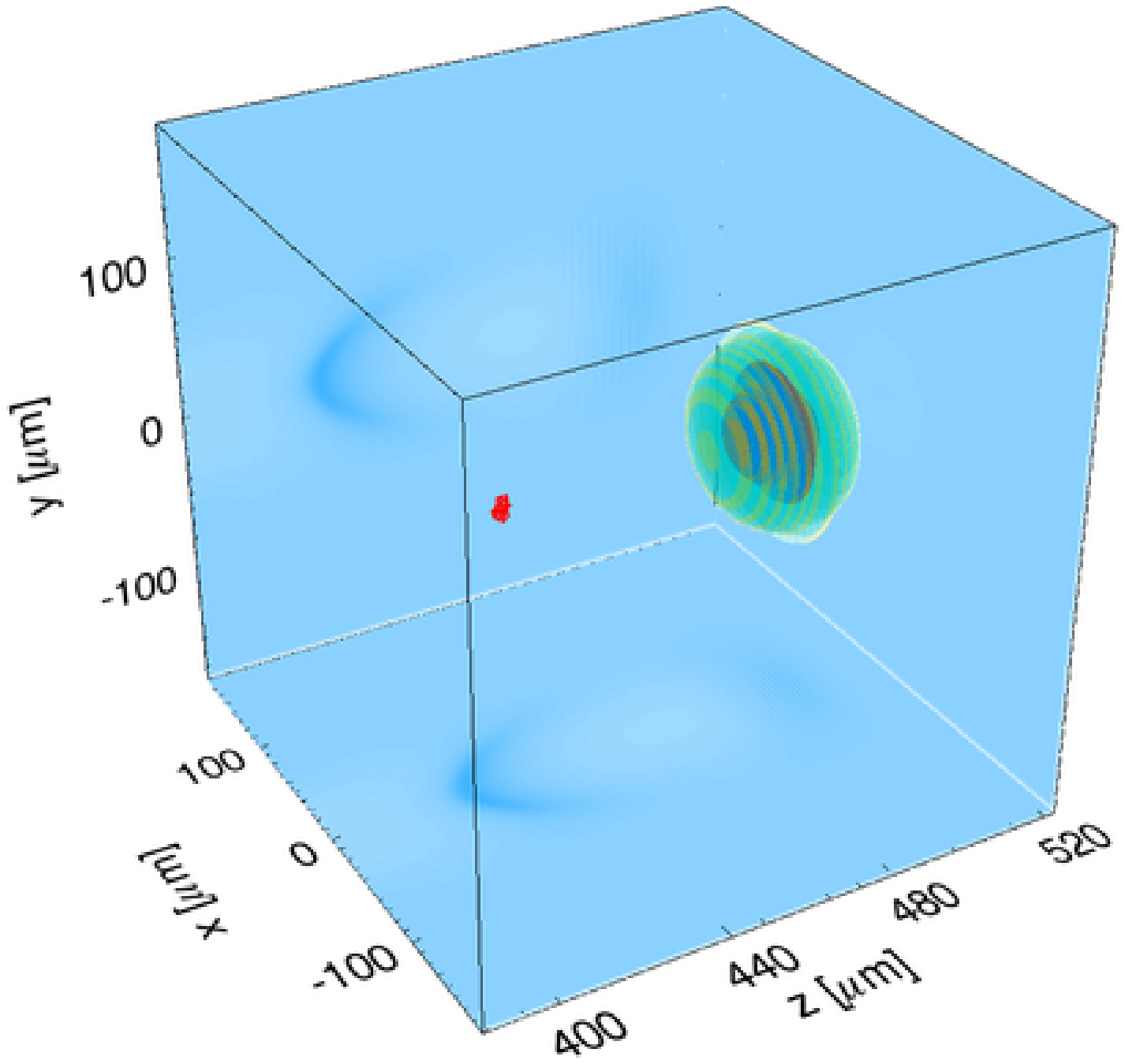}
      \put(24,78){(b)}
    \end{overpic}} &
    \subfigure{\label{fig:snapshot15}
    \begin{overpic}[width=0.21\textwidth]{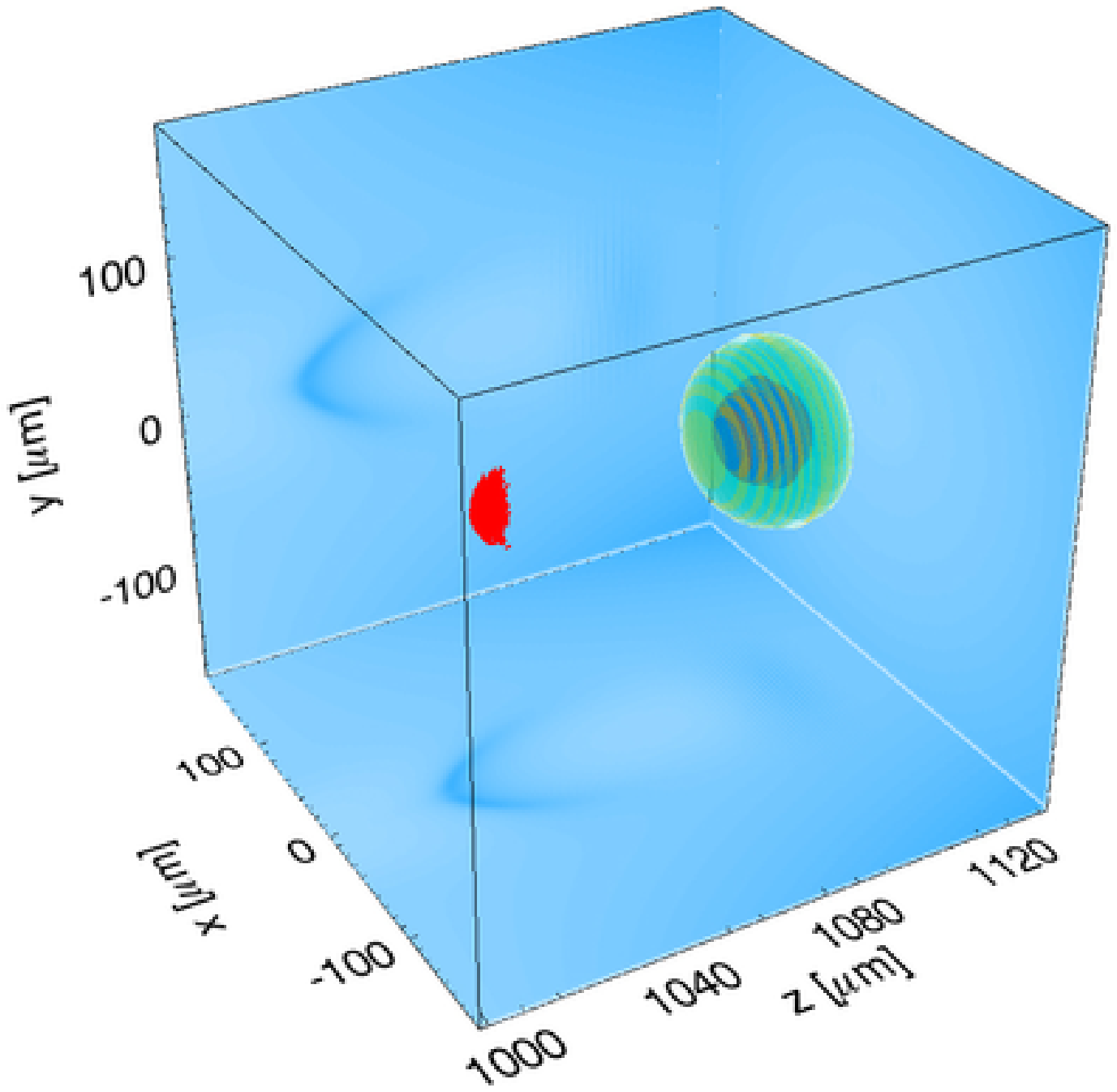}
      \put(24,78){(c)}
    \end{overpic}} &
    \subfigure{\label{fig:snapshot110}
    \begin{overpic}[width=0.21\textwidth]{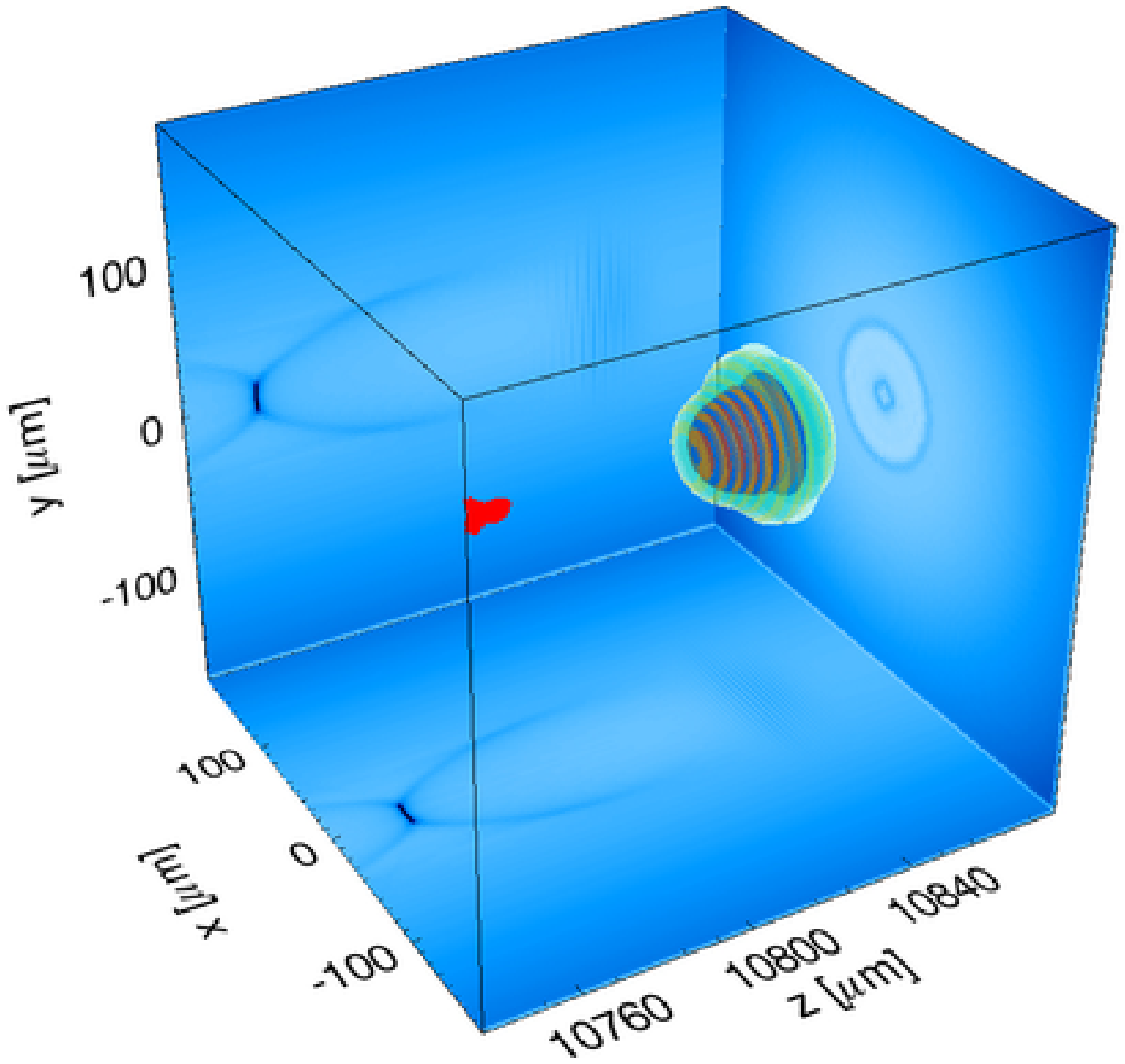}
      \put(24,78){(d)}
    \end{overpic}} \\
    \subfigure{\label{fig:laser_max_3D12}
    \begin{overpic}[width=0.23\textwidth]{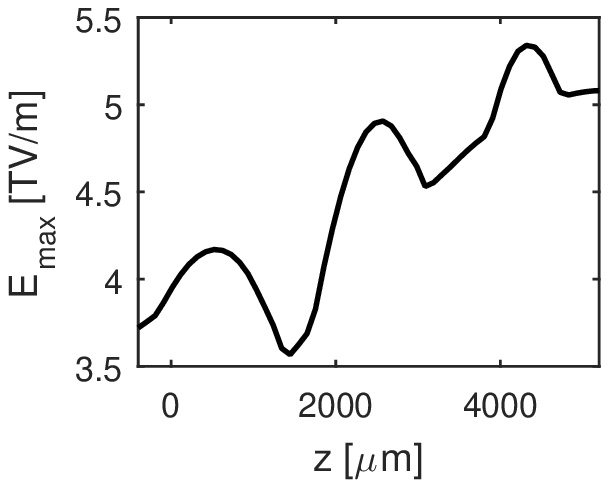}
      \put(25,61){(e)}
    \end{overpic}} &
    \subfigure{\label{fig:E_dE_vs_z}
    \begin{overpic}[width=0.22\textwidth, trim=0 1 0 -1]{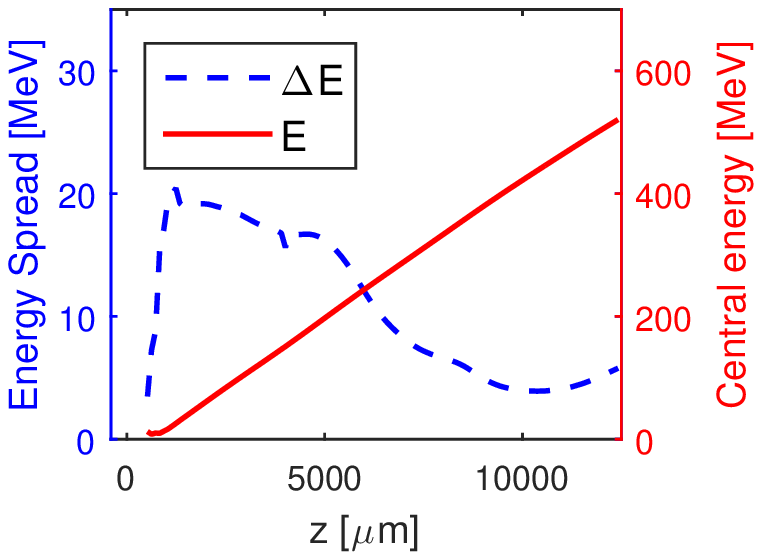}
      \put(65,64){(f)}
    \end{overpic}} &
    \subfigure{\label{fig:p1x1_3D12_15}
    \begin{overpic}[width=0.219\textwidth]{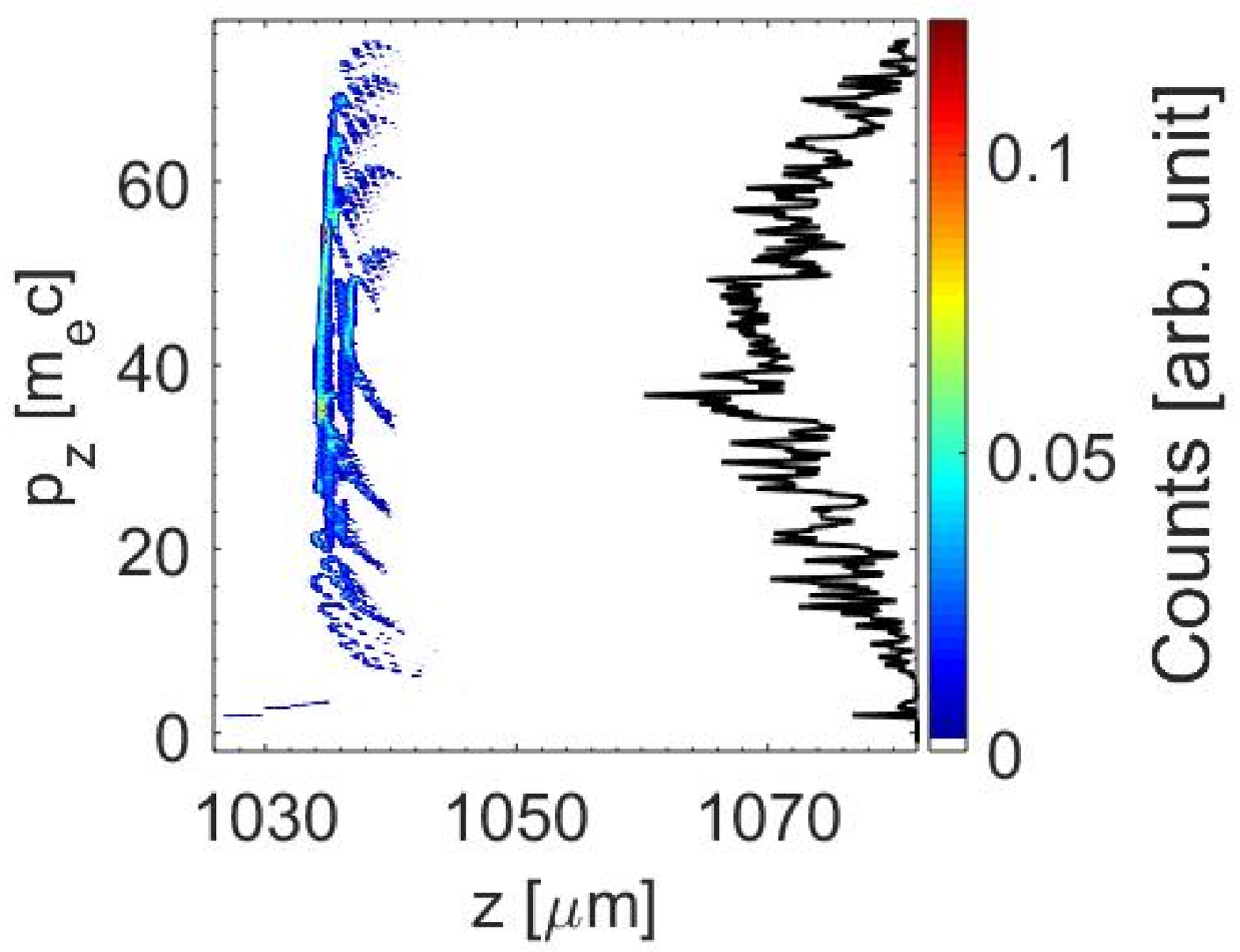}
      \put(3,65){(g)}
    \end{overpic}} &
    \subfigure{\label{fig:p1x1_3D12_110}
    \begin{overpic}[width=0.219\textwidth]{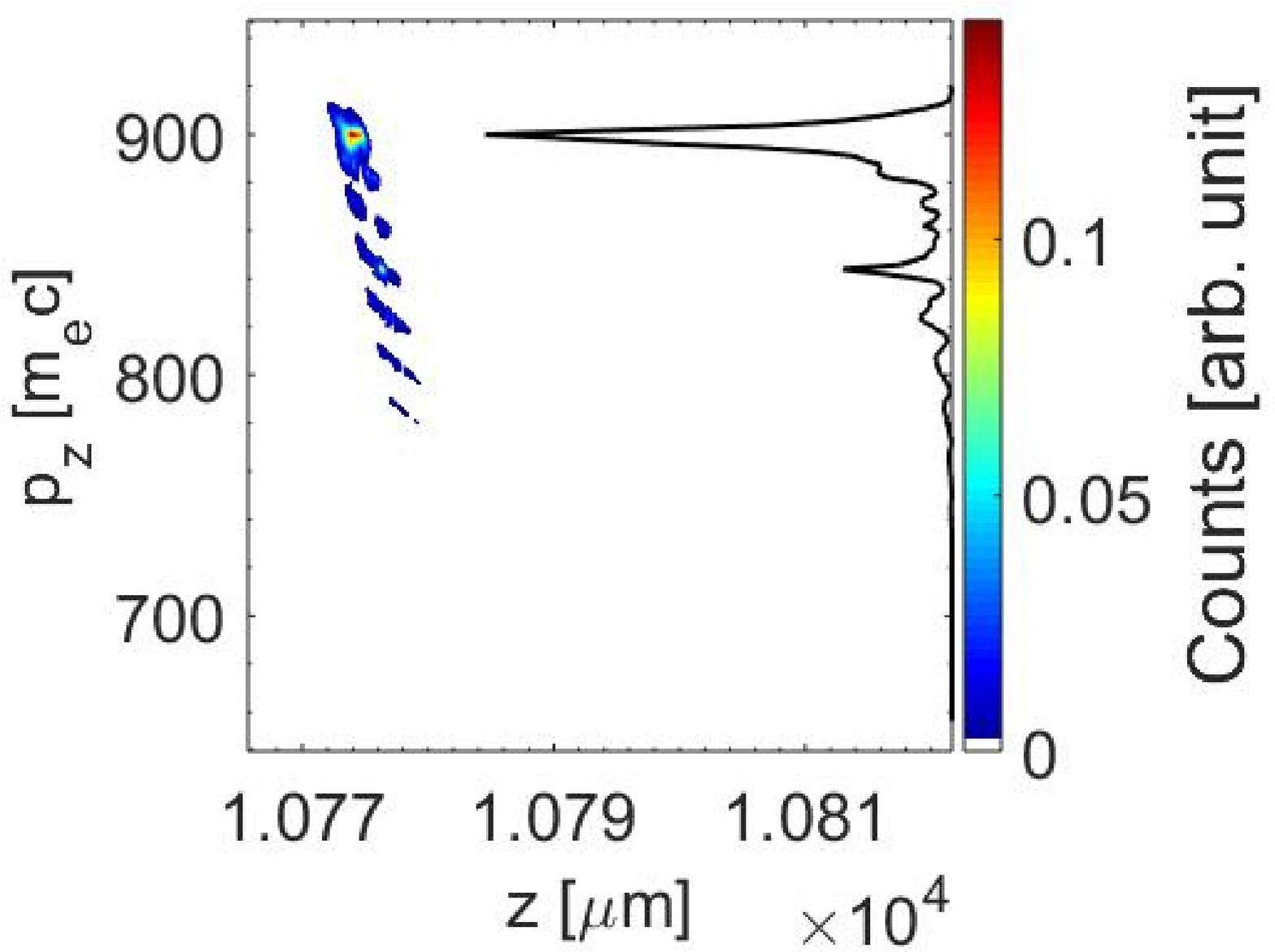}
      \put(0,65){(h)}
    \end{overpic}}
  \end{tabular}
  \caption{\label{fig:3D12} (a-d) Four snapshots of plasma electron density distributions. The blue color shows the plasma density projections to 3 planes. The contours in the right half of the boxes show the laser electric fields at different distances. The red dots in (b-d) show the injected electrons (multimedia view 1). (e) The laser peak electric field vs.\ $z$. (f) The energy and energy spread evolution of injected beam. (g) and (h) show the $(p_z, z)$ phase space distributions of injected electrons at the distances corresponding to the snapshot of (c) and (d), respectively. The black curves in (g) and (h) are the projections to the $p_z$ axis (multimedia view 2).}
\end{figure*}


We performed a series of full three dimensional (3D) particle-in-cell (PIC) simulations to study the injection and acceleration processes using the code OSIRIS~\cite{OSIRIS}. According to the previous conclusions, we choose carbon as the injection element, and both $\rm CH_4$ and $\rm CO_2$ can be the candidate. But $\rm CO_2$ has the priority because of the safety considerations. The electric field amplitude of the $2.4\ \rm \mu m$ component is chosen to be 3.07 TV/m, corresponding to the normalized vector potential of $a_{10}=2.295$ which can excite a relativistic wake to trap electrons. Consequently the $0.8\ \rm \mu m$ component has the normalized vector potential of $a_{20}=0.255$. From the above discussions it is clear that under this intensity, the oxygen atoms do not provide K-shell electrons. The two components of the SWBL have the same waist size of $W_{10}=W_{20}=60\ \rm \mu m$. The laser pulse component of $2.4\ \rm \mu m$ wavelength has the full-width-half-maximum (FWHM) duration of 100 fs, and the $0.8\ \rm \mu m$ component has the duration of 10 fs. Their profile maximums overlap initially. We use shorter duration of the $0.8\ \rm \mu m$ component to avoid ionization injection from multiple electric field peaks. The gas target profile is schematically shown in Fig.~\ref{fig:channel}. The uniform region from $z=0$ to $1000\ \rm \mu m$ is used for a stable SWBL injection. The channel from $z=1400\ \rm \mu m$ to infinite is used for a stable long distance acceleration, which has a matched channel depth~\cite{EsareyRMP2009}. The plasma density of the uniform region is $n_p = 1.92\times 10^{17}\ \rm cm^{-3}$, so that the theoretical injection interval is 1.8 mm according to Eq.~\ref{eq:period_z}. We choose three cases of $\rm C^{4+}$ densities for simulations: $n_{C^{4+}} = 0.28$, 0.56 and $1.2\times 10^{16}\ \rm cm^{-3}$, respectively. The last choice means we have to use pure $\rm CO_2$ in the injection stage, because each $\rm CO_2$ molecule provides 16 background plasma electrons, and $n_{\rm C^{4+}} / n_p= \frac{1}{16}$ is the maximum.

Firstly, we present the results from $n_{\rm C^{4+}} = 0.28 \times 10^{16}\ \rm cm^{-3}$. Some typical laser-plasma snapshots together with the $(p_z, z)$ phase space plots of the injected electrons are shown in Fig.~\ref{fig:3D12}. Figure~\ref{fig:snapshot2} to \ref{fig:snapshot110} are some snapshots of the acceleration structure. Figure~\ref{fig:laser_max_3D12} shows the laser peak electric field evolution. At the beginning, the laser evolves as the theoretical prediction for a SWBL with a period of $\Delta z = 1.8\ \rm mm$ according to Eq.~\ref{eq:period_z}. Later it undergoes the self-focusing and the focusing due to the channel for $z>1.4\ \rm mm$. It is worth noting that such self-focusing can occur even in uniform plasma without channel, and the self-focal length is $z_{\rm sf}=Z_R (P/P_c -1)^{-1/2}=2.7\ \rm mm$ by the weak relativistic assumptions~\cite{MZengPOP2014}. In our case, the injection stage, which has $\rm C^{4+}$, is in the range of $-400\ \rm \mu m < z < 1400\ \rm \mu m$. This is the reason why there is no injection due to self-focusing, and only one injection bunch exists as show in Fig.~\ref{fig:p1x1_3D12_15}. Although not explicitly showing here, it can be seen from the supplemental multimedia of Fig.~\ref{fig:p1x1_3D12_15} and \ref{fig:p1x1_3D12_110} that the injection starts from $z=220\ \rm \mu m$ and ends at $z=980\ \rm \mu m$. Figure~\ref{fig:E_dE_vs_z} shows the energy and energy spread evolution of the injected beam. One can see that the energy spread reaches a minimal at about $z=1\ \rm cm$ even though the energy still grows linearly after that. This is because that the phase rotation before dephasing is the main process for minimizing the energy spread. Figure~\ref{fig:p1x1_3D12_110} shows the phase space distribution at the optimal acceleration distance, i.~e.\ the distance where the relative energy spread has its minimal value. The energy spread is 0.88 \% in FWHM in this case. One may note that there is another small spike in the phase space projection in Fig.~\ref{fig:p1x1_3D12_110}. This spike is from the same injection period instead of another bunch.

\begin{figure}
  \centering
  \begin{tabular}{cc}
      \begin{minipage}{0.23\textwidth}
        \subfigure{\label{fig:charge_emittance_3D12}
        \begin{overpic}[width=\textwidth]{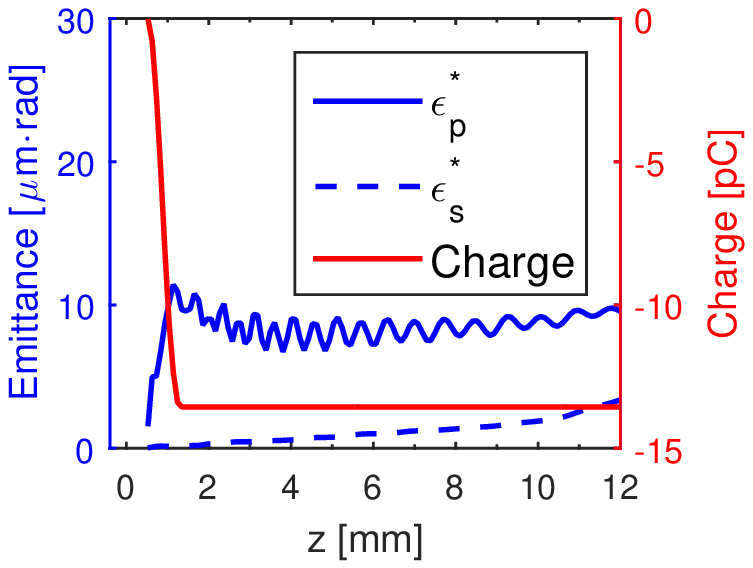}
          \put(25,65){(a)}
        \end{overpic}} \\
        \subfigure{\label{fig:charge_emittance_3D15}
        \begin{overpic}[width=\textwidth]{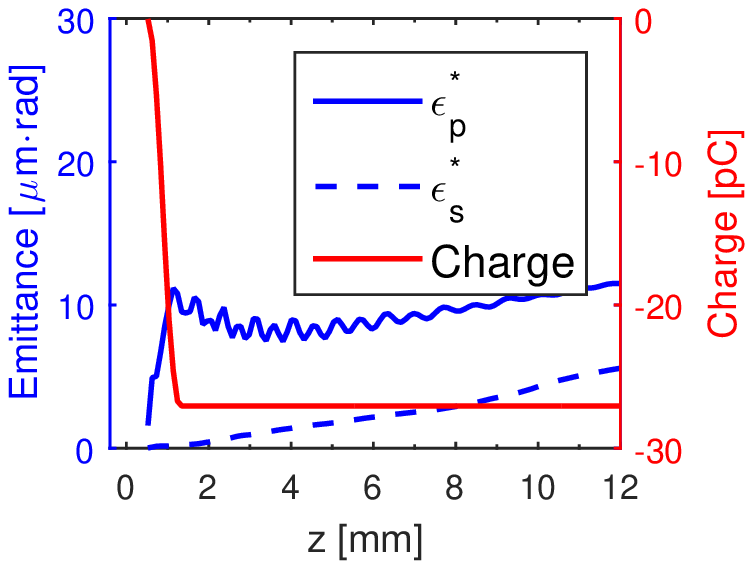}
          \put(25,65){(b)}
        \end{overpic}} \\
        \subfigure{\label{fig:charge_emittance_3D16}
        \begin{overpic}[width=\textwidth]{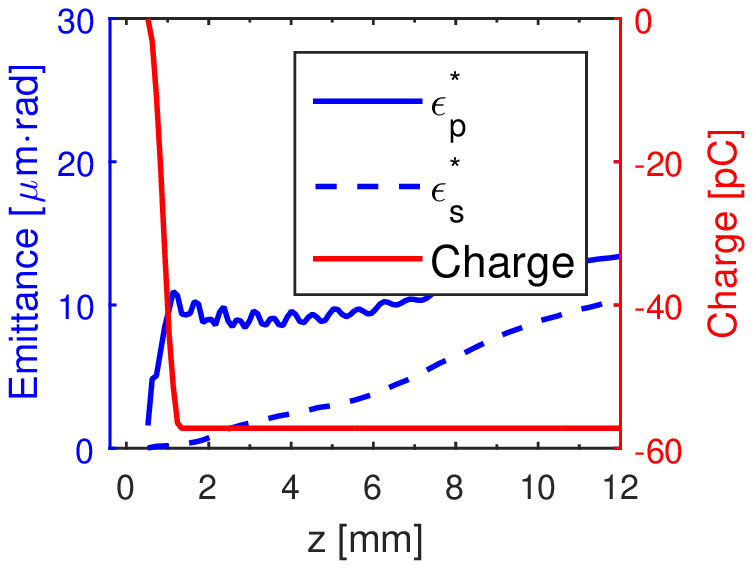}
          \put(25,65){(c)}
        \end{overpic}}
      \end{minipage} &
      \begin{minipage}{0.23\textwidth}
        \subfigure{\label{fig:C_vs_ene}
        \begin{overpic}[width=\textwidth]{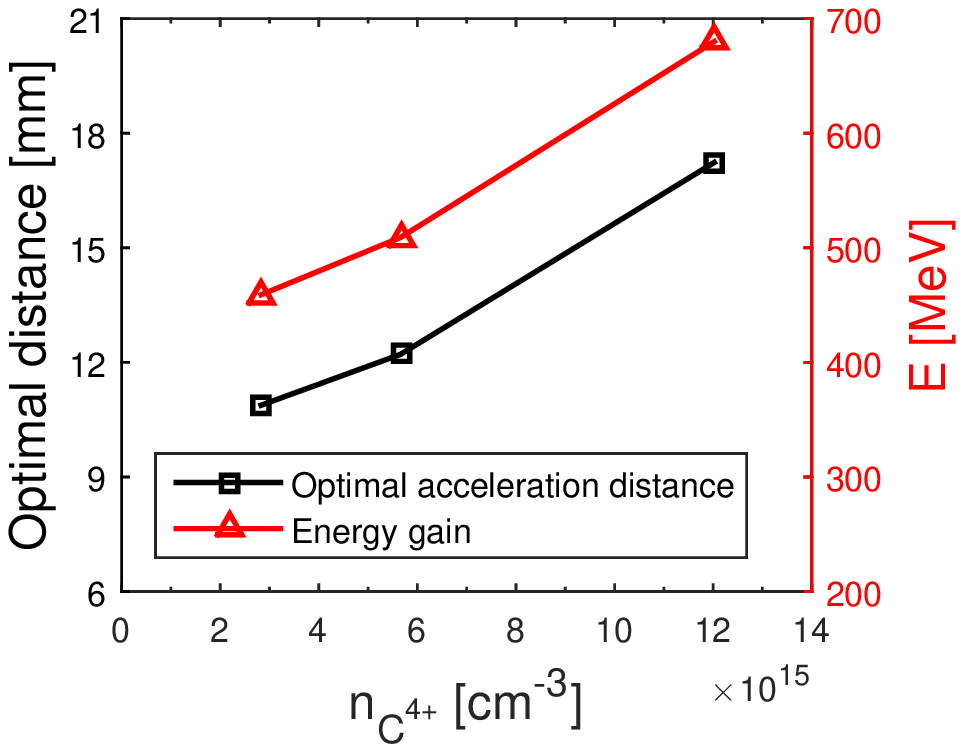}
          \put(15,65){(d)}
        \end{overpic}} \\
        \subfigure{\label{fig:C_vs_charge}
        \begin{overpic}[width=\textwidth]{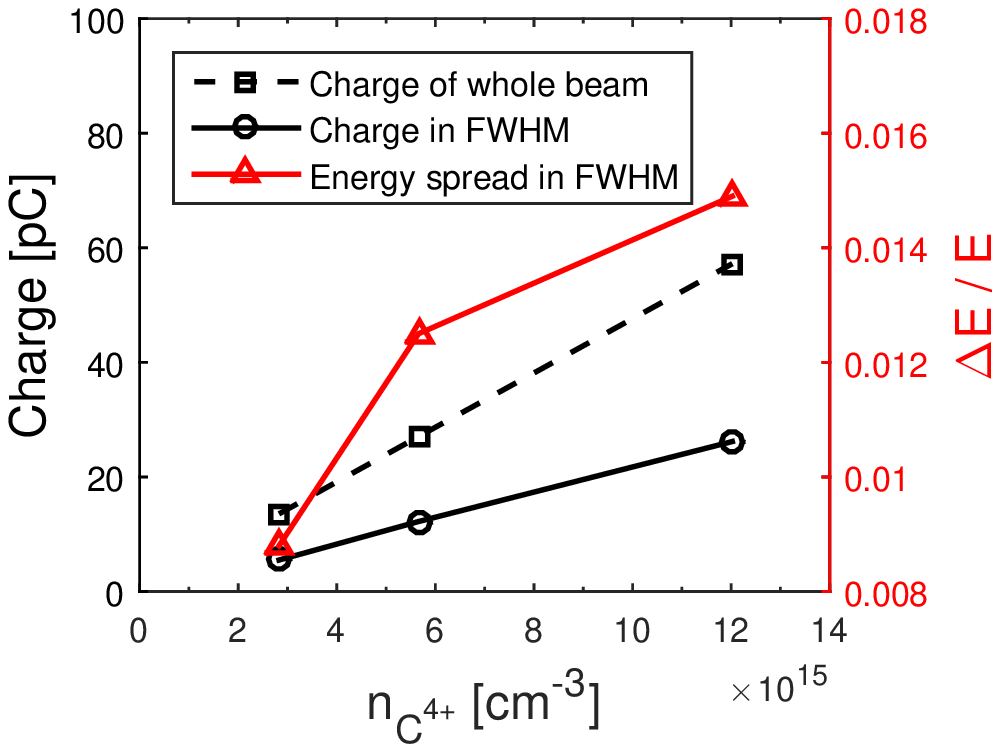}
          \put(15,20){(e)}
        \end{overpic}} \\
        \subfigure{\label{fig:spectrum}
        \begin{overpic}[width=\textwidth]{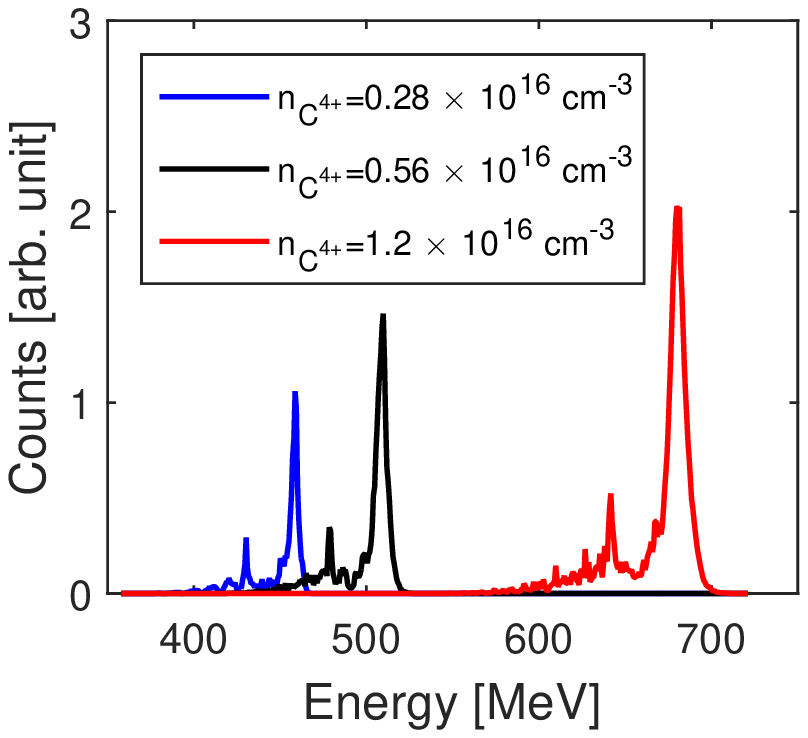}
          \put(15,20){(f)}
        \end{overpic}}
      \end{minipage}
  \end{tabular}
  \caption{\label{fig:beam_quality} Plots showing the electron beam qualities in three cases. The evolutions of the trapped electron beam charge (red curves) and the normalized transverse emittance (blue curves)  with $n_{\rm C^{4+}} = 0.28$, 0.56 and $1.2 \times 10^{16}\ \rm cm^{-3}$ are illustrated in (a), (b), and (c) respectively,  where the solid and dashed blue curves are the emittances in the directions paralleled to and perpendicular to the laser polarization direction, respectively. (d) The optimal acceleration distance and the energy gain at this distance vs.\ $n_{\rm C^{4+}}$. (e) The final charge (absolute number) of the whole beam, the charge within FWHM respect to the peak energy in the spectrum, and the relative energy spread in FWHM at optimal acceleration distance vs.\ $n_{\rm C^{4+}}$. (f) The energy spectra at the optimal acceleration distance for the three cases.}
\end{figure}

Next we present the results using different $n_{\rm C^{4+}}$: (1) $0.28 \times 10^{16}\ \rm cm^{-3}$, (2) $0.56 \times 10^{16}\ \rm cm^{-3}$ and (3) $1.2 \times 10^{16}\ \rm cm^{-3}$. The laser evolutions and the injection processes are similar, but the output electron beams show differences as one can see in Fig.~\ref{fig:beam_quality}. From Fig.~\ref{fig:charge_emittance_3D12} to \ref{fig:charge_emittance_3D16}, one can see that the beam injections only occur in the injection stage ($z<1.4\ \rm mm$), during which the emittance in the laser polarization direction $\epsilon_p^*$ also grows abruptly. In the acceleration stage ($z>1.4\ \rm mm$) $\epsilon_p^*$ oscillates and grows slowly. Similar phenomenon is also observed by others~\cite{XLXuPRSTAB2014}. One can also observe that the emittance in the other transverse direction $\epsilon_s^*$ continuously grows slowly. Figure~\ref{fig:C_vs_ene} shows the optimal acceleration distances and the energy gain at these distances for the three cases. The positive correlation of the optimal acceleration distances and $n_{\rm C^{4+}}$ is because that the beam loading effect modifies the acceleration electric field, thus a larger injected charge makes the $(p_z, z)$ phase rotation slower~\cite{TzoufrasPRL2008}. Figure~\ref{fig:C_vs_charge} shows the charges and energy spreads for the three cases. It is clear that a larger number of injected charges makes the relative energy spread larger, although the optimal acceleration distance is longer, and the corresponding energy gain is larger. We also plot the charge numbers within the FWHM in the energy spectra. One can see that about a half of the beam charge is inside the FWHM range, thus the FWHM energy spreads truly represent the beam qualities. Figure~\ref{fig:spectrum} shows the spectra of the three cases at the optimal acceleration distances.

To conclude, we have proposed a new configuration for producing high quality beams in LWFAs. The gas target has an injection stage with a uniform distribution of pure $\rm CO_2$ or $\rm CO_2$ mixed with some background gas, and an acceleration stage with transversely parabolic distributed pre-discharged plasma channels. The background gas and the gas in the acceleration stage can be a regular low-Z gas such as $\rm H_2$ or $\rm He$, and can also be $\rm O_2$ since the K-shell of $\rm O$ has extremely low ionization probability with the laser amplitude we are using. A dual-color laser pulse with a waist of $60\ \rm \mu m$ is adopted, where the main pulse has a duration of 100 fs at the wavelength of $2.4\ \rm \mu m$, and the trigger pulse has a duration of  10 fs at the wavelength of $0.8\ \rm \mu m$, but the maximums of their profile overlap. Femtosecond level temporal synchronization of these two pulses is required.  Such $2.4\ \rm \mu m$ (100 fs duration, 71 TW peak power) laser system is under design~\cite{FYWangLPL2015}, and such $0.8\ \rm \mu m$ (10 fs duration, 7.9 TW peak power) laser technique is already available~\cite{TavellaOE2006}. Output electron beam can have charge and energy spread of (13.56 pC, 0.88\%), (27.05 pC, 1.25\%) or (57.22 pC, 1.38\%) for different densities of $\rm CO_2$ cases. It is also worth noting that higher injected charge is achievable by using another injection gas which provide fewer background electrons, such as $\rm CH_4$. Our configuration can produce several times higher charge compared with other LWFA researches with energy spreads of the 1\% level. This is because the main pulse at longer wavelength and a relatively low plasma density are used in our scheme, thus with a limited power the laser pulse can have much larger waist, which can drive a larger wake structure to load a higher electron beam charge with a good quality. Our scheme is very helpful for the applications requiring high charge and low energy spread~\cite{CordeRMP2013, MChenLSA2016, CoupriePPCF2016}.

The authors would like to acknowledge the OSIRIS Consortium, consisting of UCLA and IST (Lisbon, Portugal) for the use of OSIRIS and the visXD framework. M.Z.\ appreciates the useful discussion about the laser technique with Prof.\ Lie-Jia Qian. This work is supported by the National Basic Research Program of China (2013CBA01504), the National Science Foundation of China (11421064, 11374209, 11374210), Extreme Light Infrastructure - Nuclear Physics (ELI-NP) Phase I, a project co-financed by the Romanian Government and European Union through the European Regional Development Fund, EuPRAXIA (Grant No.\ 653782), and a Leverhulme Trust Research Grant. Simulations were performed on the Supercomputer $\Pi$ at Shanghai Jiao Tong University.

%

\end{document}